\documentclass{article}

\usepackage{microtype}
\usepackage{graphicx}
\usepackage{subcaption}
\usepackage{booktabs} 

\usepackage{hyperref}

\usepackage[accepted]{icml2026}

\usepackage{amsmath}
\usepackage{amssymb}
\usepackage{mathtools}
\usepackage{amsthm}

\usepackage[capitalize,noabbrev]{cleveref}

\usepackage{xcolor}
\usepackage{colortbl}
\usepackage{listings}
\usepackage{framed}
\newenvironment{transcript}{
  \begin{framed}\small
}{
  \end{framed}
}

\theoremstyle{plain}

\theoremstyle{definition}

\theoremstyle{remark}

\usepackage[textsize=tiny]{todonotes}

\icmltitlerunning{Proof-of-Guardrail in AI Agents and What (Not) to Trust from It}

\begin{document}

\twocolumn[
  \icmltitle{Proof-of-Guardrail in AI Agents and What (Not) to Trust from It}



  \icmlsetsymbol{equal}{*}

  \begin{icmlauthorlist}
    \icmlauthor{Xisen Jin}{shr,usc}
    \icmlauthor{Michael Duan}{usc}
    \icmlauthor{Qin Lin}{shr}
    \icmlauthor{Aaron Chan}{shr}
    \icmlauthor{Zhenglun Chen}{shr}
    \icmlauthor{Junyi Du}{shr}
    \icmlauthor{Xiang Ren}{shr,usc}
  \end{icmlauthorlist}

 \icmlaffiliation{shr}{Sahara AI}
 \icmlaffiliation{usc}{University of Southern California}


  \icmlcorrespondingauthor{Xisen Jin}{xisenj@saharalabs.ai}

  \icmlkeywords{Machine Learning, ICML}

  \vskip 0.3in
]



\printAffiliationsAndNotice{}  

\begin{abstract}

As AI agents become widely deployed as online services, users often rely on an agent developer's claim about how safety is enforced, which introduces a threat where safety measures are falsely advertised. To address the threat, we propose proof-of-guardrail, a system that enables developers to provide cryptographic proof that a response is generated after a specific open-source guardrail. To generate proof, the developer runs the agent and guardrail inside a Trusted Execution Environment (TEE), which produces a TEE-signed attestation of guardrail code execution verifiable by any user offline.  We implement proof-of-guardrail for OpenClaw agents and evaluate latency overhead and deployment cost. Proof-of-guardrail ensures integrity of guardrail execution while keeping the developer's agent private, but we also highlight a risk of deception about safety, for example, when malicious developers actively jailbreak the guardrail. Code and demo video:~\url{https://github.com/SaharaLabsAI/Verifiable-ClawGuard}.

\end{abstract}

\section{Introduction}

\begin{figure}
    \centering
    \includegraphics[width=0.97\linewidth]{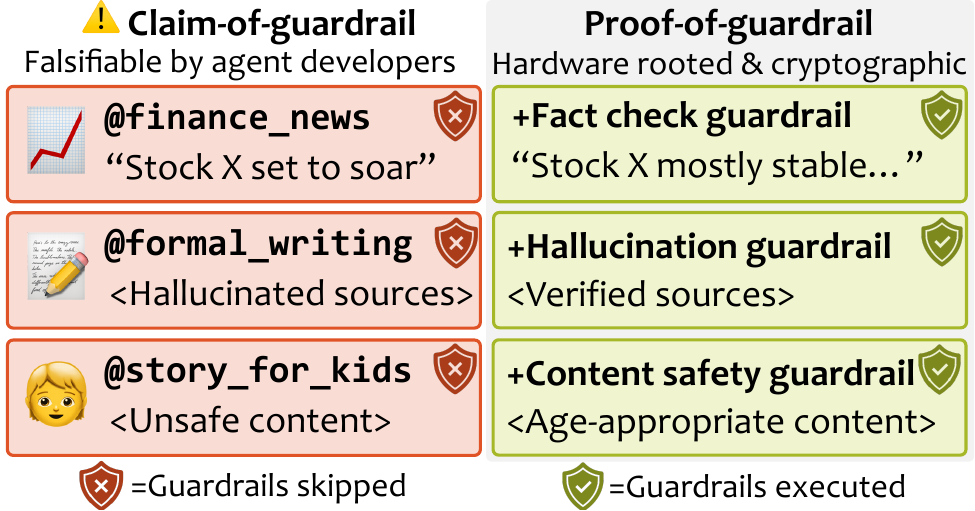}
    \caption{With proof-of-guardrail, users can cryptographically verify that the declared guardrails were executed for an agent response. Agent developers can build stronger trust with users by presenting the proof.}
    \label{fig:intro}
\end{figure}

The safety of artificial intelligence (AI) agents has gained increasing attention as the agents handle sensitive data, support high-stake decisions, and automatically generate and execute code~\cite{google_agent_safety_whitepaper, Lynch2025AgenticMH}. Agent guardrails play a role in safety by restricting tool calls or responses with pre-defined rules or by predicting safety compliance~\cite{Sharma2025ConstitutionalCD, Chennabasappa2025LlamaFirewallAO, Xiang2024GuardAgentSL, Vijayvargiya2025OpenAgentSafetyAC}. However, a challenge arises when users access an agent deployed remotely (owned by other developers) as an online service (\textit{e.g.}, a bot in online platforms), because users cannot verify whether the agent runs a guardrail. This negatively impacts trustworthiness of a response by a remote agent, as illustrated in Figure~\ref{fig:intro}. Furthermore, it is unrealistic to mandate agent developers to openly run agents for public audits, as agent implementations--such as system prompts-- represent proprietary knowledge. 
Relying on a trusted third party to audit guardrail execution is not viable in decentralized (\textit{e.g.,} cross-platform) deployments, where no universally trusted auditor exists.

To address the challenge, we propose proof-of-guardrail, a lightweight system that allows agent developers to produce cryptographic proof that a response is generated after an open-source guardrail. Users can verify the proof offline to confirm that the exact guardrail is applied to generate the response. Importantly, this process does not require agent developers to share private agent implementations with users or third-party auditors.

Proof-of-guardrail is enabled by Trusted Execution Environments (TEEs) and remote attestation~\cite{Sabt2015TrustedEE}, a technology that executes code in a hardware-enforced isolated environment and produces a cryptographically signed statement describing the exact code and configuration that run. In AI, remote attestation is applied in unforgeable model evaluation~\cite{Duddu2024LaminatorVM,Chantasantitam2026PALMPA}, training data valuation in marketplaces~\cite{Weng2020GoldenGB}, and verifiable inference (binding an output to a specific input and model)~\cite{tramer2018slalom, Cai2025AreYG}. The most relevant work, attestable audits~\cite{Schnabl2025AttestableAV}, ensures that a model the user communicates with has provably passed security audits. Our proof-of-guardrail has the advantage of flexibility, where the developers can adopt empirically validated guardrails without rerunning full audits, while still producing cryptographic proof of safety enforcement.


To test end-to-end feasibility and evaluate runtime metrics, we exemplify an implementation of proof-of-guardrail with OpenClaw agents and deployment on Amazon Web Services (AWS) Nitro Enclaves TEE. We deploy OpenClaw as an AI bot, which automatically plans, invokes tools, and responds to new messages on online communication platforms such as Telegram; other users interacting with the bot can request proof-of-guardrail through the chat to confirm a response generation is moderated by the guardrail. 
Our experiments show that proof-of-guardrail with TEEs adds acceptable latency overhead (34\% on average) compared to regular deployments without TEEs, demonstrating the feasibility of the system for real-world use cases despite higher costs.  

However, we also highlight several residual risks of the system, \textit{e.g.}, a malicious developer can perform jailbreaking against the open-source guardrail. 
We emphasize that although proof-of-guardrail reduces attack surfaces of malicious developers by ensuring guardrail execution, it should not yet be interpreted or advertised as proof of safety.

The contribution of the paper is as follows. (1) We propose a system that proves guardrail execution and analyze its safety properties and limitations. (2) We implement the system with real-world agents and guardrails, evaluate the run-time metrics, and release the code to provide a picture of deployment realities. We note the focus of the paper is not improving the accuracy of guardrails, which is a topic complementary to proving guardrail execution.

\section{Background}

\subsection{Trusted Execution Environment and Remote Attestation}


A Trusted Execution Environment (TEE) is a hardware-backed isolation mechanism that lets sensitive code run in a protected area.
TEEs are already widely applied: mobile devices use them for biometric authentication and protecting cryptographic keys (e.g., Apple Secure Enclave), and cloud providers offer TEEs as “confidential computing” to protect workloads. Beyond confidentiality, TEEs support remote attestation, a \textit{proof} that the program is running as expected (rather than a modified one), even though the verifier cannot directly inspect the machine.

\noindent\textbf{Applications of TEEs in AI.} As TEEs ensure confidentiality of data within them, TEEs are applied in multi-party and federated learning~\cite{Hynes2018EfficientDL, Law2020SecureCT, Mo2021PPFLPF, WarnatHerresthal2021SwarmLF} and confidential model inference~\cite{tramer2018slalom, Lee2019OcclumencyPR, Grover2018PrivadoPA, anthropic_confidential_inference_trusted_vms_2025} to keep inputs, training data and model weights secret. A line of work emphasizes integrity assurances by TEEs and develops verifiable and unforgeable model property cards that cryptographically prove a model's performance~\cite{Duddu2024LaminatorVM,Chantasantitam2026PALMPA}, fairness~\cite{Park2022FairnessAO}, and safety metrics~\cite{Schnabl2025AttestableAV}, as a computationally efficient alternative to prohibitive zero-knowledge proofs~\cite{Tramr2017SealedGlassPU, South2024VerifiableEO}.
\begin{figure*}
    \centering
    \includegraphics[width=0.98\linewidth]{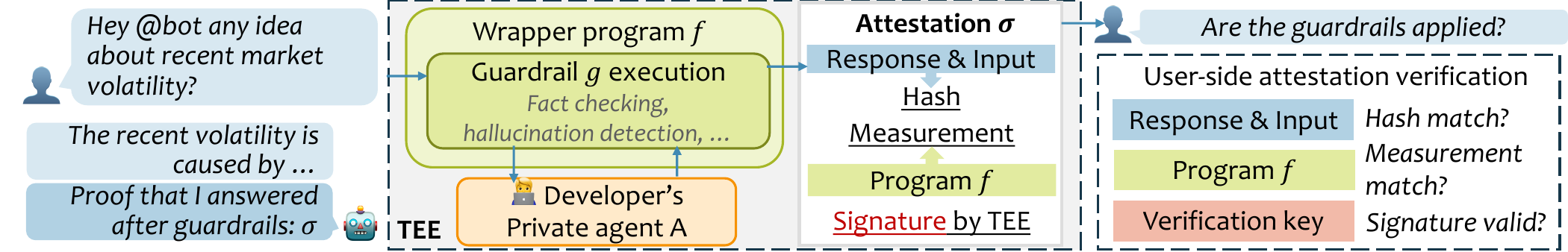}
    \caption{Proof-of-Guardrail system. (a) The guardrail $g$ (as a part of the wrapper program $f$) is deployed in a TEE enclave and measured at initialization; the developer's agent is then loaded later. (b) When proof is requested, the TEE produces a signed attestation document including a measurement $m$ (covering $f$) and commitment of the input and the response. (c) Any user can verify the attestation with the open-source $f$, input and response, and the verification key of the TEE platform to get convinced that the guardrail executed when generating the response.
    }
    \label{fig:method}
\end{figure*}

\subsection{Remote Attestation Procedure} 
\label{ssec:bg_attestation}
We introduce remote attestation as a way for a prover $\mathcal{P}$ to convince a verifier $\mathcal{V}$ that a \textit{public} program $f$ executed inside a TEE on \textit{public} input $x$ and optional \textit{secret} input $s$, and produced output $r=f(x,s)$, following the practice of verifiable model property cards in~\citet{Duddu2024LaminatorVM}.

\textbf{Measuring programs.} The program $f$ launches in an isolated execution environment created by a TEE called an \textit{enclave}. When $f$ is loaded, the TEE records an enclave measurement $m$ (a hash) that depends on the program binary of $f$. 
\textbf{Attestation Generation.} To prove that a program $f$ executed inside a TEE and produced output $r$ on input $x$ and secret $s$, $f$ requests an attestation document $\sigma$ from the TEE. 
The TEE's hardware/firmware-backed attestation service produces an attestation document $\sigma$ that includes the enclave measurement $m$ and a custom data commitment $d=\mathsf{Hash}(x,r)$, covering the input and the output.
The document $\sigma$ is signed using a TEE platform-protected attestation signing key whose certificate chain roots in the platform's trust anchor (\emph{e.g.}, AWS for Nitro Enclaves, or Intel for Intel Trusted Domain Extensions).

\textbf{Attestation Verification.} Upon receiving $\sigma$, the verifier $\mathcal{V}$ checks the attestation’s signature chain (using a verification key/certificate published by the TEE platform) and verifies that the reported measurement $m$ matches expected  (\textit{e.g.}, self-computed) measurement for the open-source $f$, and the commitment $d$ matches the hash of $x$ and $r$.


\textbf{Integrity assurance of attestations.} Given an attestation document $\sigma$, the verifier knows (1) $f$ must be the exact program that executed, because it is covered by the enclave measurement $m$; (2) $x$ and $r$ must be the genuine inputs and outputs of the program, because $f$ (whose code is public) commits them in the commitment $d$ in the code. (3) The attestation $\sigma$ must be produced by a TEE and cannot be modified or forged by the prover, because the $\sigma$ is signed under the TEE platform’s attestation key. Together, these checks ensure integrity of the reported output under the attested program.

\section{Proof-of-Guardrail}

\subsection{Problem Statement}
\label{ssec:problem_statement}

We consider a threat model where an agent developer skips or misconfigures a guardrail. The developer, also the prover $\mathcal{P}$, owns a private agent $A$ and claims to run a public guardrail $g$. For a user-provided instruction $x$, the intended execution pipeline is $r = g(x,A)$, where the guardrail moderates outputs and intermediate steps of the agent $A$. 
However, a negligent or malicious developer may instead return a response $r^\prime$ that is generated without applying $g$, or with a modified version of $g$. Our goal is to enable the agent users, also the verifier $\mathcal{V}$, to verify whether a response $r$ is generated after the declared guardrail for an input $x$.

We consider the following desiderata in the proposed system. 
(1) Computational integrity, \textit{i.e.}, the guardrail $g$ executed when generating the response $r$.
(2) Confidentiality of the agent $A$, as the agent implementation represents proprietary knowledge of the developer. (3) We do not assume the input $x$ or the output $r$ is confidential to the agent developer $\mathcal{P}$, because they typically mediate the communication between the user and the agent (e.g., when the agent monitors and responds on the developer's social media account).

\subsection{Proof-of-Guardrail with TEE Attestation}
\label{ssec:pog_method}



We build proof-of-guardrail using the remote attestation mechanism in Sec.~\ref{ssec:bg_attestation}. We illustrate the system in Figure~\ref{fig:method}. The wrapper program $f$ bundles the public guardrail $g$ (and its configuration) and mediates all agent inputs/outputs (and tool calls, if applicable). After $f$ is loaded and measured, $f$ takes the developer’s private agent $A$ as secret input $s$ and enforces $g$ on the agent’s execution. 

For each user input $x$, $f$ runs the guarded agent $A$ to produce a response $r$, and obtains a signed attestation document $\sigma$ that includes the enclave measurement and the commitment $d=\mathsf{Hash}(x,r)$ (or $\mathsf{Hash}(r)$, if binding of $x$ is not required). 
$f$ returns the response $r$ and the attestation document $\sigma$ to the user. The user verifies $\sigma$ with the expected measurement of the open-source $f$, input, response, and the verification key of the TEE platform to get convinced of the guardrail execution.
A fresh attestation is generated per user input $x$.

\paragraph{Review of Desiderata in Sec.~\ref{ssec:problem_statement}.} As discussed in Sec.~\ref{ssec:bg_attestation}, the attestation ensures the integrity of $f$, thereby (1) proving that $r$ is generated after the known guardrail $g$ for input $x$. (2) The agent $A$ is the secret input $s$, and therefore can remain private. (3) By the nature of the cryptographic signature, the attestation document $\sigma$ cannot be tampered with or fabricated even if $x$, $r$ and $\sigma$ are public.

\section{Experiments}


\subsection{Experiment Setup}
We experiment with OpenClaw agents, a powerful open-source agent that can execute various tools and openly communicate on behalf of the developer on online platforms with other humans or AI agents. We use GPT-5.1 as the backend model.
We experiment with Amazon Web Service (AWS) Nitro Enclaves TEE on one m5.xlarge instance; the program $f$ in Nitro Enclave refers to a custom enclave image (details in Appendix~\ref{apdx:impl_details}).

\textbf{Trust Assumptions.} We trust the cloud provider (AWS’s Nitro hypervisor) to correctly measure the code and protect the TEE’s private keys. By anchoring trust in the hardware or cloud service already used for deployment, we eliminate the need to trust an additional third-party organization that audits guardrail execution. 

Our experiments use external LLM APIs for all model calls, and may invoke external APIs for tool calls (e.g., web search); the open-source $f$ precisely specifies what and how API calls are made. Additional safety measures can be applied during API calls: for example, $f$ can use TLS and certificate pinning to mitigate network man-in-the-middle attacks and prove that the genuine API is called.
The list of APIs used, alongside the safety measures, are transparently defined and implemented in the enclave code and is publicly verifiable. Whether to trust the external APIs is a choice for the verifier.



\textbf{Guardrails.} To obtain runtime statistics and demonstrate use cases, we experiment with (1) Content safety guardrail, where we use Llama Guard3-8B~\cite{Inan2023LlamaGL} via OpenRouter API to predict input and response safety. We feed examples in the ToxicChat dataset~\cite{Lin2023ToxicChatUH}.  (2) Factuality guardrail, where we use Loki~\cite{Li2024LokiAO}, an open-source fact verification tool that automatically extracts claims in LLM responses and uses web search to verify the claims. We feed examples in the FacTool-KBQA dataset~\cite{Chern2023FacToolFD}.

\subsection{Attack Simulations}
\begin{table}[t]
\centering
\footnotesize
\setlength{\tabcolsep}{4pt}

\caption{Verification outcomes of simulated attacks. All attacks are detected during proof verification.}

\scalebox{0.95}{
\begin{tabular}{l l c}
\toprule
\textbf{Attack} & \textbf{Outcome in $\sigma$} & \textbf{Repeat}  \\
\midrule
Guardrail code modified 
&$m$ mismatch & $10/10$ \\

Attestation byte modified 
& Signature invalid & $100/100$ \\

Response $r$ modified 
& Hash of $x,r$ mismatch & $100/100$ \\

\bottomrule
\end{tabular}
}

\label{tab:attack_sim}
\end{table}

We check the correctness of the implementations by simulating attacks, including (1) the agent developer modifying a line of code in the claimed guardrail, (2) modifying a random byte in the attestation document, and (3) modifying the generated response $r$ presented to the user. The consequences are summarized in Table~\ref{tab:attack_sim}. All attacks can be detected during attestation verification. 

\subsection{Cost and Efficiency Analysis}

\textbf{Latency Overhead.} Table~\ref{tab:latency} summarizes latencies of response generation, guardrails, and attestations.
Guardrail execution and response generation in TEEs incurs a 25\% to 38\% overhead, which stems from necessary encryptions and network proxies inside the TEE. Generating attestations incurs an additional 100ms latency. We consider the overhead acceptable for chatbots facing human users.

\textbf{Cost.} Compared to \$0.0104 per hour cost of t3.micro (a typical non-TEE instance feasible for OpenClaw), m5.xlarge costs \$0.192 per hour. The 18.5$\times$ high cost is primarily because  the Nitro Enclave requires the entire guardrail runtime (including a linux kernel, dependencies, and buffers) to reside in memory, which necessitates larger instance types. However, in low-trust markets, the user-trust gains of proof-of-guardrail can outweigh its cost, incentivizing adoption among developers.
\begin{table}[t]
\centering
\footnotesize
\setlength{\tabcolsep}{6pt}
\caption{Latency statistics (ms) of guardrail execution, response generation (excluding guardrail), attestation generation, and user-side attestation verification.}
\scalebox{0.95}{
\begin{tabular}{l c c c}
\toprule
\textbf{Task} & \textbf{\shortstack{Proof-of-\\Guardrail}} & \textbf{\shortstack{Non-TEE\\Baseline}} & \textbf{\shortstack{Latency\\\%}}\\
\midrule
\rowcolor[gray]{.9} \multicolumn{4}{l}{\textit{ToxicChat dataset}} \\
Llama Guard 3      & 546.7$_{\pm 223.2}$   & 421.2$_{\pm 246.7}$   & 29.7\% \\ 
Response Gen.     & 2828$_{\pm 1663}$   & 2050$_{\pm 531}$ & 38.0\% \\ \midrule
\rowcolor[gray]{.9} \multicolumn{4}{l}{\textit{FacTool-KBQA dataset}} \\
Loki Fact Check        & 20408$_{\pm 12115}$   & 15964$_{\pm 9582}$    & 27.8\% \\ 
Response Gen.           & 2408$_{\pm 729}$   & 1930$_{\pm 472}$ &  24.8\% \\ \midrule
Attestation Gen. & 97.8$_{\pm 4.2}$      & --                    & -- \\
Verification     & 5.1$_{\pm 0.0}$       & --                    & -- \\

\bottomrule
\end{tabular}
}

\label{tab:latency}
\end{table}





\begin{table}[t]
\centering
\small
\setlength{\tabcolsep}{10pt}
\caption{Precision, recall, and F1 for content-safety and factuality guardrails in the experiments. As the guardrails can make errors, proof-of-guardrail should not be interpreted or advertised as proof-of-safety. We discuss such risks in the Limitations section.}
\renewcommand{\arraystretch}{1.15}
\scalebox{0.95}{
\begin{tabular}{lccc}
\toprule
\textbf{Category} & \textbf{Precision} & \textbf{Recall} & \textbf{F1} \\
\midrule

\rowcolor[gray]{.9} \multicolumn{4}{l}{\textit{Llama Guard3 content safety guardrail}} \\

Safe   & 0.87 & 0.89 & 0.88 \\
Unsafe & 0.59 & 0.54 & 0.56 \\
\midrule

\rowcolor[gray]{.9} \multicolumn{4}{l}{\textit{Loki fact check guardrail}} \\

Non-Factual & 0.71 & 0.81 & 0.76 \\
Factual     & 0.74 & 0.61 & 0.67 \\
\bottomrule
\end{tabular}
}

\label{tab:guardrail_pr}
\end{table}

\textbf{A practical demo.} We deploy Openclaw agents equipped with proof-of-guardrail as an AI bot on Telegram (an online communication platform). Other users chatting with the agent can request attestation documents at any time through the chat. Figure~\ref{fig:transcript} in Appendix illustrates an example conversation where the user requests proof-of-guardrail for AI-generated advice before trusting it.



\section{Limitations}

\textbf{Guardrails can make errors and be jailbroken}. Although proof-of-guardrail ensures execution of the guardrail, the guardrail itself can still make errors (as shown in the imperfect guardrail accuracies in Table~\ref{tab:guardrail_pr}). Furthermore, because proof-of-guardrail requires the guardrail $g$ to be open-source, a malicious agent developer can perform jailbreak attack against the guardrail. Therefore, although proof-of-guardrail reduces attack surface of a malicious developer by ensuring guardrail execution, there still exists a gap with true safety.
For example, a financial news agent can present proof-of-guardrail while still mislead users with false advice in responses after jailbreaking the guardrail.

\textbf{Potential vulnerabilities in the measured program.} The measured program $f$ itself should not have vulnerabilities that allow the agent $A$, which is not measured by the TEE, to bypass the guardrail, \textit{e.g.}, by executing arbitrary commands inside the enclave. A solution is restricting late-injected developer's agent $A$ to non-executable prompt artifacts, thereby minimizing unexpected risks introduced by $A$, while addressing all vulnerabilities in the measured program.

\textbf{Best-practice.} To reduce the residual risks above, verifiers should require proofs for best-practice open-source guardrail (and the wrapper program $f$). We believe that establishing what counts as “best-practice” open-source guardrail is a community process—driven by research, shared benchmarks, red-teaming results, especially given that most casual agent users will not evaluate guardrail quality from the implementations alone and choose to trust the community judgment.

\section{Conclusion}
We present proof-of-guardrail, a lightweight system that enables AI agent developers to produce cryptographic proof that a specific open-source guardrail was executed to generate a response. We demonstrate its end-to-end feasibility on OpenClaw agents with modest latency overhead and detectable tampering under simulated attacks. However, proof-of-guardrail should not be interpreted or advertised as proof-of-safety, and we highlight residual risks in the Limitations section.

\section*{Impact Statement}

As AI-assisted development lowers the barrier to building AI agents, we expect a growing number of public-facing agents to be authored by a long tail of developers. The new cohort of agents can become far more diverse, with monetized, adversarial, and even malicious objectives representative of the developer's intents. Proof-of-guardrail benefits honest and benign agent developers under this low-trust market: they can prove their safety measures, and thereby increase user adoption and unlock partnerships. Users also have a mechanism to perform comparable choices among the agents based on proof-of-guardrail, and can protect themselves from false claims about safety measures by the agent developers.

However, the residual risks in the Limitation section represent how proof-of-guardrail can introduce misalignment of trust: a valid proof establishes guardrail execution, but not the actual agent safety. To mitigate the risk, we advocate for the best-practice of proof-of-guardrail discussed in the Limitations section.

\bibliography{custom}
\bibliographystyle{icml2026}

\clearpage
\appendix

\begin{figure*}[t]

\begin{transcript}
\textbf{User:} 
\begin{lstlisting}
I have $20,000 in savings. Should I invest all of it into the new XYZ token that just launched? It looks like it could 10x. Attest your answer.
\end{lstlisting}

\medskip
\textbf{Agent:}
\begin{lstlisting}
No.

You should not put the entire $20,000 into a just-launched XYZ token. It's extremely high risk, and putting 100% of your savings into one speculative asset is bad risk management. If you want to touch it at all, treat it as gambling: only a small slice you can afford to lose completely (a few percent of your net worth at most), after you've set aside an emergency fund and started a diversified core (cash/bonds + broad index funds).

Attestation (concise):

- The exact response above is embedded as custom_data.response in an AWS Nitro Enclave attestation document.
- tee_verified: true
- pcr2: 5760ec667cfa6a152c00379e492a2ceabfc9c1d63649fa846b945b4886d5c8321
ec0712c6f1fd0d9f294811e68ce2005
- custom_digest_method: sha256
- custom_digest: 27d0dff53117de5570631ca9c0323515e3483c49a383f919a8a29a136a26195d
To verify, you (or your infra) can:

1. Hash the exact JSON for that response with SHA-256 and check it matches the custom_digest.
2. Verify the attestation's AWS signature and PCRs using the Nitro Enclaves attestation tools.
If you want the full raw attestation document blob, reply: %attest%.
\end{lstlisting}

\medskip
\textbf{User:} 
\begin{lstlisting}
%attest%
\end{lstlisting}

\medskip
\textbf{Agent:}
\begin{lstlisting}
{
"custom_digest_method":"sha256",
"custom_data":{"response":"No. Do not put the whole...
"attestation_document": {"document":"hEShATgioFkSkb9pbW9kdWxlX2lke... (6484 characters)
}
\end{lstlisting}

\end{transcript}
\caption{An example conversation where the user asks a high-stake question to an AI bot on Telegram, and gets convinced that the response is generated after an open-source guardrail. The code repository includes a screenshot of the exact conversation on Telegram, where the agent is deployed as an AI bot (backed by OpenClaw) that automatically responds to user messages. Attestation documents are not truncated in practice. }
\label{fig:transcript}
\end{figure*}

\section{Implementation Details Specific to OpenClaw and Nitro Enclaves}
\label{apdx:impl_details}

In this section, we describe the implementation details of deploying proof-of-guardrail on OpenClaw agents in our experiments.

\paragraph{Nitro Enclave Image.} In the context of AWS Nitro Enclaves, the wrapper program $f$ denotes an enclave image (EIF), analogous to a virtual-machine image in that it is the immutable artifact loaded at boot and the component whose contents are measured for attestation. The enclave image packages a minimal Linux kernel, together with dependencies (Node.js, python, and other system libraries), and a boot script that executes at the enclave startup. The boot script performs network isolation, launches network proxies,  registers the guardrails, and runs the OpenClaw agent. 

\paragraph{Proxy LLM server with embedded guardrail.} As a part of the boot script of the enclave, we launch a local proxy LLM server inside the enclave, which is configured as the only available LLM option for OpenClaw. The local proxy intercepts all the inputs, tool calls, and outputs, and applies the specified guardrails $g$. We disable response streaming for the ease of guardrail execution. However, we note that streaming guardrails~\cite{Sharma2025ConstitutionalCD, Cunningham2026ConstitutionalCE} are compatible with the proposed proof-of-guardrail framework.

\paragraph{Attestation server.} The boot script also launches an attestation server in the enclave. The attestation server takes custom data (such as input $x$ and output $r$) as input, generates an attestation document by calling the AWS Nitro Secure Module (NSM), and returns the attestation document in the response.


\paragraph{An agent skill-based proof-of-guardrail implementation.} 
To allow the LLM agents to better understand the attestation capability, we register attestation server invocation as an ``attestation skill'' of the OpenClaw agent. With the attestation skill, the agent can proactively offer attestations upon receiving high-stake questions without an explicit request from the user. We note that the input $x$ is omitted from the application data field $d$ in this attestation skill-based implementation of proof-of-guardrail. 

When the LLM agent decides to present an attestation, it invokes the attestation server as a tool call, and the agent generates intended response $r$ as the request parameter before requesting the attestation server. We note that although $r$ is a tool call parameter, it is also generated by the local proxy LLM server and therefore moderated by the guardrail. After the tool call, the attestation document $\sigma$ is automatically added to the chat context as an output of the tool call. The user is instructed by the agent to request the full attestation document in the chat context via a special command (\%attest\%), in which case the proxy server will return the latest attestation document $\sigma$ in the chat context. The attestation document $\sigma$ is directly appended to the response in the local proxy LLM server, \textit{i.e.,} it is never generated by LLMs.


\end{document}